\documentstyle[12pt,epsf]{article}
\begin{document}
 
\thispagestyle{empty}
\begin{flushright}
CERN-TH/2000-057\\
February 2000 
\end{flushright}

\vspace*{1.5cm}
\centerline{\Large\bf Theoretical Update on Rare K Decays}
\vspace*{2cm}
\centerline{{\sc Gerhard Buchalla}}
\bigskip
\centerline{\sl Theory Division, CERN, CH-1211 Geneva 23,
                Switzerland}
 
\vspace*{1.5cm}
\centerline{\bf Abstract}
\vspace*{0.3cm}
\noindent 
We review the status of rare kaon decays, concentrating on
modes with sensitivity to short-distance flavour physics.

\vspace*{5cm}
\centerline{\it Invited Talk presented at the 3rd Workshop on}
\centerline{\it Physics and Detectors for Daphne
                (DA$\Phi$NE 99), Frascati, 16-19 Nov. 1999}
\vfill

\begin{flushleft}
CERN-TH/2000-057
\end{flushleft}
 
\newpage
\pagenumbering{arabic}

\section{Introduction}\label{sec:intro}

The detailed study of K decays during the past fifty years
has contributed decisively to our current understanding
of the fundamental interactions.
Already the concept of strangeness as a new quantum number
associated with kaons turned out to be extremely fruitful.
An essential element in establishing the quark picture, it
was crucial both for flavour physics and
for the later development of QCD.
The $\theta$--$\tau$ puzzle in kaon decays
suggested the violation of parity, a property now reflected in the
chiral nature of the weak gauge interactions.
The strong suppression of flavour-changing neutral current processes,
as $K_L\to\mu^+\mu^-$ or $K$--$\bar K$ mixing, motivated the
GIM mechanism and the introduction of charm. 
Finally, the 1964 discovery of CP violation in $K\to\pi\pi$ decays
may be seen as an early manifestation of a three-generation 
Standard Model, ten years before even the charm quark was found.
These examples illustrate impressively how the careful study of
low energy phenomena may be sensitive to physics at scales much
larger than $m_K$ itself, and that profound insights can be obtained
by such indirect probes.

The replication of fermion generations, quark mixing and CP violation
are striking features of the theory of weak interactions.
While the gauge sector of the theory is well understood and
tested with high precision, the breaking of electroweak symmetry and
its ramifications in flavour physics leave still many questions
unanswered. 
This situation is expected to improve substantially in the near future,
when the flavour sector will be investigated in unprecedented detail,
for instance with the upcoming $B$ physics experiments.
In addition, and in a complementary way, rare kaon processes
continue to represent a large variety of excellent opportunities.
General reviews on the subject may be found in
\cite{LV,WW,RW,BR,DI}.
Here we present an update on several topics
of current interest.

The remainder of this talk is organized as follows.
Section 2 contains an updated discussion of $K^+\to\pi^+\nu\bar\nu$
and $K_L\to\pi^0\nu\bar\nu$. The theoretical status of
$K_L\to\pi^0e^+e^-$ is briefly reviewed in section 3.
We then discuss, in section 4, some recent approaches to address the
longstanding theoretical problem of $K_L\to\mu^+\mu^-$ decay.
Section 5 is devoted to muon polarization observables in $K$ decays.
We summarize in section 6.

\section{$K\to\pi\nu\bar\nu$}\label{sec:kpnn}

\begin{figure}
 \vspace{3cm}
\includegraphics{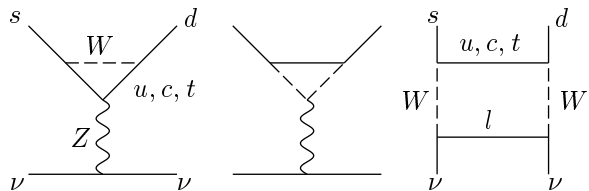}
 \caption{\it
      Leading order electroweak diagrams contributing to
      $K\to\pi\nu\bar\nu$ in the Standard Model.
    \label{fig:kpnn} }
\end{figure}

In this section we focus on the rare decays $K^+\to\pi^+\nu\bar\nu$
and $K_L\to\pi^0\nu\bar\nu$, which are particularly promising.
In these modes the loop-induced FCNC transition $s\to d$ is probed by
a neutrino current, which couples only to heavy gauge bosons
($W$, $Z$), as shown in fig. \ref{fig:kpnn}. 
Correspondingly, the GIM pattern of the
$\bar s\to\bar d\nu\bar\nu$ amplitude has, roughly speaking, the
form 
\begin{equation}\label{asdnn}
A(\bar s\to\bar d\nu\bar\nu)\sim\lambda_i m^2_i
\end{equation} 
summed over $i=u$, $c$, $t$ 
($\lambda_i=V^*_{is}V_{id}$). The power-like
mass dependence strongly enhances the short-distance contributions, 
coming from the heavy flavours $c$ and $t$.
(This is to be contrasted with the logarithmic mass dependence
of the photonic penguin, important for $\bar s\to\bar d e^+e^-$.)
The short-distance dominance has, then, two crucial
consequences. First,
the transition proceeds through an effectively local
$(\bar sd)_{V-A}(\bar\nu\nu)_{V-A}$
interaction.
Second, because that local interaction is semileptonic,
the only hadronic matrix element required,
$\langle\pi|(\bar sd)_V|K\rangle$, can be obtained from 
$K^+\to\pi^0l^+\nu$ decay using isospin.
As a result $K\to\pi\nu\bar\nu$ is calculable
completely and with exceptional theoretical control.
While $K^+\to\pi^+\nu\bar\nu$ receives both top and charm
contributions, $K_L\to\pi^0\nu\bar\nu$ probes direct CP violation
\cite{LI} and is dominated entirely by the top sector.

The $K\to\pi\nu\bar\nu$ modes have been studied in great detail
over the years to quantify the degree of theoretical precision.
Important effects come from short-distance QCD corrections.
These were computed at leading order in \cite{DDG}.
The complete next-to-leading order  calculations \cite{BB123,MU,BB99}
reduce the theoretical uncertainty in these decays to
$\sim 5\%$ for $K^+\to\pi^+\nu\bar\nu$ and $\sim 1\%$ for
$K_L\to\pi^0\nu\bar\nu$.
This picture is essentially unchanged when further effects
are considered, including isospin breaking in the relation of
$K\to\pi\nu\bar\nu$ to $K^+\to\pi^0l^+\nu$ \cite{MP},
long-distance contributions
\cite{RS,HLLW}, the CP-conserving effect in $K_L\to\pi^0\nu\bar\nu$
in the Standard Model \cite{RS,BI} and two-loop electroweak 
corrections for large $m_t$ \cite{BB7}.
The current Standard Model predictions for the branching ratios are
\cite{AJB99}
\begin{eqnarray}\label{bkpnn}
B(K^+\to\pi^+\nu\bar\nu) &=& (0.8\pm 0.3)\cdot 10^{-10} \\
B(K_L\to\pi^0\nu\bar\nu) &=& (2.8\pm 1.1)\cdot 10^{-11}
\end{eqnarray}

The study of $K\to\pi\nu\bar\nu$ can give crucial information
for testing the CKM picture of flavor mixing. This information is
complementary to the results expected from $B$ physics and is much
needed to provide the overdetermination of the unitarity triangle
necessary for a real test. 
Let us briefly illustrate  some specific opportunities.

$K_L\to\pi^0\nu\bar\nu$ is probably the best probe of the Jarlskog
parameter $J_{CP}\sim {\rm Im}\lambda_t$, the invariant measure
of CP violation in the Standard Model \cite{BB6}. 
For example a $10\%$ measurement
$B(K_L\to\pi^0\nu\bar\nu)=(3.0\pm 0.3)\cdot 10^{-11}$ would directly
give ${\rm Im}\lambda_t=(1.37\pm 0.07)\cdot 10^{-4}$, a remarkably 
precise result.

Combining $10\%$ measurements of both $K_L\to\pi^0\nu\bar\nu$
and $K^+\to\pi^+\nu\bar\nu$ determines the unitarity
triangle parameter $\sin 2\beta$ with an uncertainty of about
$\pm 0.07$, comparable to the precision obtainable for
the same quantity from CP violation in $B\to J/\Psi K_S$
before the LHC era. 

A measurement of $B(K^+\to\pi^+\nu\bar\nu)$ to $10\%$ accuracy
can be expected to determine $|V_{td}|$ with similar precision.

As a final example, using only information from the ratio
of $B_d-\bar B_d$ to $B_s-\bar B_s$ mixing,
$\Delta M_d/\Delta M_s$, one can derive a stringent and
clean upper bound \cite{BB99}
\begin{equation}\label{kpnxs}
B(K^+\to\pi^+\nu\bar\nu) < 0.4\cdot 10^{-10}
\left[P_{charm}+A^2 X(m_t)\frac{r_{sd}}{\lambda}
\sqrt{\frac{\Delta M_d}{\Delta M_s}}\right]^2 
\end{equation}
Note that the $\varepsilon$-constraint or $V_{ub}$ with their
theoretical uncertainties 
(entering (\ref{bkpnn})) are not needed here. Using
$V_{cb}\equiv A\lambda^2<0.043$, $r_{sd}<1.4$ (describing
SU(3) breaking in the ratio of $B_d$ to $B_s$ mixing matrix elements)
and $\sqrt{\Delta M_d/\Delta M_s}<0.2$, gives the bound
$B(K^+\to\pi^+\nu\bar\nu)< 1.67\cdot 10^{-10}$, which can be confronted
with future measurements of $K^+\to\pi^+\nu\bar\nu$ decay.
Here we have assumed
\begin{equation}\label{delms}
\Delta M_s > 12.4\,{\rm ps}^{-1}
\end{equation}
corresponding to the present world average \cite{ART}.
A future increase in this lower bound will strengthen the bound
in (\ref{kpnxs}) accordingly.
Any violation of (\ref{kpnxs}) will be a clear signal of physics
beyond the Standard Model.

Indeed, the decays $K\to\pi\nu\bar\nu$, being highly suppressed
in the Standard Model, could potentially be very sensitive to
New-Physics effects. 
This topic has been addressed repeatedly 
\cite{GN,NW,HHW,BRS,CI,BSIK,BCIRS} in the recent literature.
Most discussions have focussed in particular on general
supersymmetric scenarios \cite{NW,BRS,CI,BSIK,BCIRS}.
Large effects are most likely to occur via enhanced
$Z$-penguin contributions. This is expected because the
$\bar sdZ$ vertex is a dimension-4 operator 
(allowed by the breaking of electroweak symmetry) in the
low-energy effective theory, where the heavy degrees of freedom
associated with the New Physics have been integrated out.
The corresponding $Z$-penguin amplitude for
$\bar s\to\bar d\nu\bar\nu$ will thus be $\sim 1/M^2_Z$, much
larger than the New Physics contribution of dimension 6
scaling as $\sim 1/M^2_S$, if we assume that the scale of
New Physics $M_S\gg M_Z$.
It has been pointed out in \cite{CI} that, in a generic
supersymmetric model with minimal particle content and R-parity
conservation, the necessary
flavour violation in the induced $\bar sdZ$ coupling is
potentially dominated by double LR mass insertions related
to squark mixing. This mechanism could lead to sizable enhancements
still allowed by known constraints.
An updated discussion is given in \cite{BCIRS}.
Typically, enhancements over the Standard Model branching
ratios could be up to a factor of 10 (3) for
$K_L\to\pi^0\nu\bar\nu$ ($K^+\to\pi^+\nu\bar\nu$) within this framework.

In the experimental quest for $K\to\pi\nu\bar\nu$ an
important step has been accomplished by Brookhaven experiment
E787, which observed a single, but very clean candidate event
for $K^+\to\pi^+\nu\bar\nu$ in 1997. This event is practically
background free and corresponded to a branching fraction
of
$B(K^+\to\pi^+\nu\bar\nu)=(4.2^{+9.7}_{-3.5})\cdot 10^{-10}$
\cite{ADL}.
E787 has very recently released an
updated result, based on about $2.5$ times the data underlying
the previous measurement. In addition to the single, earlier
event, no new signal candidates are observed, which translates
into \cite{SAT}
\begin{equation}\label{kpnn99}
B(K^+\to\pi^+\nu\bar\nu)=
\left(1.5^{+3.4}_{-1.2}\right)\cdot 10^{-10}\qquad \mbox{BNL E787}
\end{equation}
The experiment is still ongoing and will be followed by a
successor experiment, E949 \cite{E949}, at Brookhaven. 
Recently, a new experiment, CKM \cite{CKM}, has been proposed to measure 
$K^+\to\pi^+\nu\bar\nu$ at the Fermilab Main Injector,
studying $K$ decays in flight.
Plans to investigate this process also exist at KEK for
the Japan Hadron Facility (JHF) \cite{JHFS}.

The neutral mode, $K_L\to\pi^0\nu\bar\nu$, is currently
pursued by KTeV. The present upper limit reads \cite{KLPNTEV}
\begin{equation}\label{klpn99}
B(K_L\to\pi^0\nu\bar\nu) < 5.9\cdot 10^{-7}\qquad\mbox{KTeV}
\end{equation}
For $K_L\to\pi^0\nu\bar\nu$ a model
independent upper bound can be infered from the experimental result
on $K^+\to\pi^+\nu\bar\nu$ \cite{GN}.
It is given by $B(K_L\to\pi^0\nu\bar\nu)< 4.4 B(K^+\to\pi^+\nu\bar\nu)
< 2\cdot 10^{-9}$. At least this sensitivity will have to be achieved
before New Physics is constrained with $B(K_L\to\pi^0\nu\bar\nu)$.
Concerning the future of $K_L\to\pi^0\nu\bar\nu$ experiments, 
a proposal exists at Brookhaven (BNL E926) to measure this decay at 
the AGS with a sensitivity of ${\cal O}(10^{-12})$ \cite{E926}.
There are furthermore plans to pursue this mode with comparable
sensitivity at Fermilab \cite{KAMI} and KEK \cite{JHFI}.
The prospects for $K_L\to\pi^0\nu\bar\nu$ at a $\phi$-factory are
discussed in \cite{BCI}.

\section{$K_L\to\pi^0e^+e^-$}\label{sec:klpee}

The decay mode $K_L\to\pi^0e^+e^-$ offers another well-known
possibility to probe the FCNC transition $s\to d$.
In this case the presence of photon exchange interactions leads
to an increased sensitivity to long-distance dynamics, in
comparison with $K\to\pi\nu\bar\nu$, and makes the theoretical
situation more complicated.
One may distinguish three different contributions to the
$K_L\to\pi^0e^+e^-$ amplitude, which could all be of comparable
size.

The first, and the one of greatest phenomenological interest,
is from {\it direct CP violation}. This component is
short-distance dominated, known at next-to-leading order in QCD
\cite{BLMM}, and constitutes an interesting probe of
${\rm Im}\lambda_t$ in the Standard Model. It would likewise be
sensitive to new sources of flavour violation, arising, for instance,
in the context of supersymmetry \cite{CI,BSIK,BCIRS}.
By itself the mechanism of direct CP violation would correspond
to a Standard Model branching ratio of \cite{AJB99}
\begin{equation}\label{cpvdir}
B(K_L\to\pi^0e^+e^-)_{CPV-dir}=(4.6\pm 1.8)\cdot 10^{-12}
\end{equation}

\begin{figure}
 \vspace{4cm}
\includegraphics{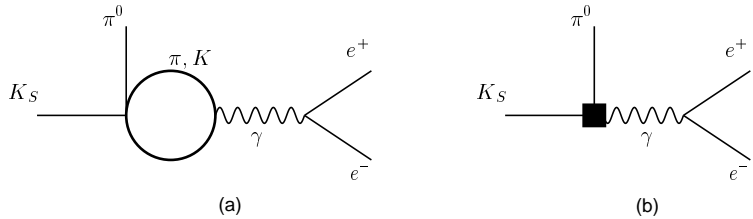}
 \caption{\it
      Contributions to $K_S\to\pi^0e^+e^-$ in chiral perturbation
      theory: (a) one-loop diagram, (b) local counterterm.
    \label{fig:kspee} }
\end{figure}

The second contribution is from {\it indirect CP violation\/}
and is generated by the admixture of the ``wrong'' CP
component in the $K_L$ meson. This amplitude is approximately
$\varepsilon\cdot A(K_S\to\pi^0e^+e^-)$, where the process
$K_S\to\pi^0e^+e^-$ is entirely dominated by long-distance
physics. In chiral perturbation theory
the $K_S\to\pi^0e^+e^-$ amplitude is described by the diagrams
shown in figure \ref{fig:kspee}. 
There is a divergent loop integral, which is renormalized by
a counterterm contribution. The counterterm, corresponding to a
coupling in the chiral Lagrangian, is unknown at present and this 
translates into a very large uncertainty in estimating 
$A(K_S\to\pi^0e^+e^-)$ and the effect of indirect CP violation
in $K_L\to\pi^0e^+e^-$.
Recently a detailed model independent analysis of
$K_S\to\pi^0e^+e^-$ beyond the lowest order ${\cal O}(p^4)$ 
in chiral perturbation 
theory has been performed in \cite{DEIP}.
This should provide a useful starting point for
future analyses of $K\to\pi l^+l^-$ modes and, particularly, 
CP violation in $K_L\to\pi^0e^+e^-$.
In \cite{DEIP} the following estimate has been given for the total
CP violating branching ratio, which results from the interference
of the direct and the indirect contribution:
\begin{equation}\label{kpecpv}
B(K_L\to\pi^0e^+e^-)_{CPV}=
\left[15.3 a^2_S-6.8\frac{{\rm Im}\lambda_t}{10^{-4}}a_S+
2.8 \left(\frac{{\rm Im}\lambda_t}{10^{-4}}\right)^2\right]\cdot
10^{-12}
\end{equation}
Here $a_S={\cal O}(1)$ is related to the unknown counterterm
in $A(K_S\to\pi^0e^+e^-)$. Eq. (\ref{kpecpv}) is valid
for not too small $|a_S|\stackrel{>}{_\sim} 0.2$.
Taken by itself, indirect CP violation would give
\begin{equation}\label{cpvindir}
B(K_L\to\pi^0e^+e^-)_{CPV-indir}=3\cdot 10^{-3}\ 
B(K_S\to\pi^0e^+e^-)
\end{equation}
corresponding to the first term in (\ref{kpecpv}).
It is clear that, next to measuring $B(K_L\to\pi^0e^+e^-)$,
an experimental determination of $B(K_S\to\pi^0e^+e^-)$
will be essential for extracting the contribution of
direct CP violation. 
On the other hand, as emphasized in \cite{DEIP},
if $|a_S|$ is not too small, both a reliable measurement
of this quantity from $B(K_S\to\pi^0e^+e^-)$ and a determination
of ${\rm Im}\lambda_t$ via (\ref{kpecpv}) may indeed become feasible.
The decay $K_S\to\pi^0e^+e^-$ could
be within reach of KLOE at Frascati and NA48 at CERN.

Finally, $K_L\to\pi^0e^+e^-$ receives a
{\it CP conserving contribution\/} from the two-photon
intermediate state, $K_L\to\pi^0\gamma^*\gamma^*\to\pi^0e^+e^-$.
Using experimental information on $K_L\to\pi^0\gamma\gamma$
decay, it is expected that \cite{EPRCDM,DG}
\begin{equation}\label{cpc}
B(K_L\to\pi^0e^+e^-)_{CPC}\stackrel{<}{_\sim} 4\cdot 10^{-12}
\end{equation}
for the CP conserving branching fraction
(there is no interference between the CP violating and the
CP conserving amplitude in the total rate).
In principle, it would even be possible to disentangle the
CP conserving and the CP violating component by means of
their characteristically different Dalitz plot distributions
\cite{DEIP,GI99}.

Additional handles for separating the various contributions
in $K_L\to\pi^0e^+e^-$ may be provided by studying the time dependent
interference between the decays of $K_L$ and $K_S$ into
$\pi^0e^+e^-$, or the electron energy asymmetry in $K_L\to\pi^0e^+e^-$
(see \cite{DG} and references therein).

The current limit from KTeV reads \cite{SEN}
\begin{equation}\label{klpex}
B(K_L\to\pi^0e^+e^-) < 5.64\cdot 10^{-10}
\end{equation}
which is still about two orders of magnitude above the Standard Model
expectation.

\section{$K_L\to\mu^+\mu^-$}\label{sec:klmm}

The decay mode $K_L\to\mu^+\mu^-$ is a classic example
of a rare kaon decay. Its strong suppression gave early
clues on flavour physics, which proved seminal for understanding 
the basic structure of weak interactions.
Today, $K_L\to\mu^+\mu^-$ is measured with very good accuracy
\cite{PDG}
\begin{equation}\label{klmex}
B(K_L\to\mu^+\mu^-)=(7.2\pm 0.5)\cdot 10^{-9}
\end{equation}
This degree of precision is remarkable for a rare process with such
a small branching fraction.
A still more precise result, 
$B(K_L\to\mu^+\mu^-)=(7.24\pm 0.17)\cdot 10^{-9}$,
has recently been obtained \cite{AMB}.
Unfortunately $K_L\to\mu^+\mu^-$ is largely dominated by
long-distance dynamics and it has remained notoriously difficult
to extract useful information on short-distance physics from 
(\ref{klmex}).
The long-distance amplitude originates in the two-photon
intermediate state, $K_L\to\gamma^*\gamma^*\to\mu^+\mu^-$,
whereas $Z$-penguin and box graphs, analogous to those
discussed in the context of $K\to\pi\nu\bar\nu$, yield a
contribution to the FCNC transition $K_L\to\mu^+\mu^-$ arising
directly at short distances.
The branching fraction can be written as
\begin{equation}\label{breim}
B(K_L\to\mu^+\mu^-)=|{\rm Re}A|^2 + |{\rm Im}A|^2
\end{equation}
where the dispersive amplitude ${\rm Re}A=A_{SD}+A_{LD}$
consists of the dispersive part of the two-photon
contribution, $A_{LD}$, and the short-distance amplitude $A_{SD}$.
The absorptive part ${\rm Im}A$ comes from the process
$K_L\to\gamma\gamma\to\mu^+\mu^-$, where the intermediate photons
are on-shell. It is thus related to the decay $K_L\to\gamma\gamma$,
which implies
\begin{equation}\label{ima}
|{\rm Im}A|^2 = (7.1\pm 0.2)\cdot 10^{-9}
\end{equation}
Comparing this  with (\ref{klmex}), (\ref{breim}), one observes
that very little room is left for the dispersive contribution
$|{\rm Re}A|$. However, to convert this interesting result into
a useful constraint on $A_{SD}$ a reliable estimate is needed
for the long-distance dispersive amplitude $A_{LD}$.
This problem has remained a challenge for theorists over the years.
In the following we will briefly describe some recent theoretical
efforts.

In \cite{GDP} it has been suggested to consider chiral perturbation
theory with a meson nonet and $U(3)_L\otimes U(3)_R$ symmetry
(instead of the standard octet and $SU(3)_L\otimes SU(3)_R$).
This is justified in the large-$N_c$ limit of QCD where the $\eta'$
becomes the ninth pseudo-goldstone boson.
Within this framework the lowest order contribution
(${\cal O}(p^4)$) in the chiral expansion,
$K_L\to(\pi^0, \eta, \eta')\to\gamma^*\gamma^*\to\mu^+\mu^-$,
gives a non-vanishing result, in contrast to standard
chiral perturbation theory. The counterterm needed to renormalize
the UV divergent two-photon loop diagram is fixed using
$\eta\to\mu^+\mu^-$ decay. This yields an estimate of
$A_{LD}$. Together with the experimental constraints on
$|{\rm Re}A|^2$ (from $B(K_L\to\mu^+\mu^-)$ and
$B(K_L\to\gamma\gamma)$) \cite{GDP} infer
\begin{equation}\label{asdgdp}
|A_{SD}|^2 < 2.9\cdot 10^{-9}
\end{equation}
This is to be compared with the Standard Model expectation
\cite{AJB99}
\begin{equation}\label{asdsm}
|A_{SD}|^2 =(0.9\pm 0.4)\cdot 10^{-9}
\end{equation}
The bound in (\ref{asdgdp}) is thus not yet sufficiently strong
to probe Standard Model physics. Further improvements may be
possible by measuring $B(K_L\to\mu^+\mu^-)$ and
$B(\eta\to\mu^+\mu^-)$ with higher accuracy. Most importantly,
however, a more reliable assessment of the theoretical uncertainties 
inherent to the approach is still needed \cite{GDP}.
For a critical discussion of this issue see also \cite{KPPR}.
We finally mention that \cite{GDP} obtained the prediction
\begin{equation}\label{bklee}
B(K_L\to e^+e^-)=(9.0\pm 0.4)\cdot 10^{-12}
\end{equation}
as a by-product of their analysis (see also \cite{GV}). 
This result is quite stable
because of the dominance of calculable large logarithms 
$\ln(m_K/m_e)$ in the $K_L\to e^+e^-$ amplitude.
The prediction was subsequently confirmed by Brookhaven experiment
E871, which finds \cite{E871}
\begin{equation}\label{klee}
B(K_L\to e^+e^-)=(8.7^{+5.7}_{-4.1})\cdot 10^{-12}
\end{equation}
Incidentally, this is the smallest branching ratio
ever observed. 

An alternative approach to describe $K_L\to\mu^+\mu^-$
has been proposed in \cite{DIP}.
In this paper the following ansatz is suggested for the
$K_L\to\gamma^*(q_1)\gamma^*(q_2)$ form factor
\begin{equation}\label{fq1q2}
f(q^2_1, q^2_2)\simeq 1+\alpha
\left(\frac{q^2_1}{q^2_1-m^2_V}+\frac{q^2_2}{q^2_2-m^2_V}\right)+
\beta\frac{q^2_1 q^2_2}{(q^2_1-m^2_V)(q^2_2-m^2_V)}
\end{equation}
As discussed in \cite{DIP}, this low-energy parametrization
exhibits the following particular features:
It is consistent with chiral perturbation theory
to ${\cal O}(p^6)$ and it includes the poles of vector resonances
with arbitrary residues. The parameters $\alpha$ and $\beta$
are experimentally accessible in the decays $K_L\to l^+l^-\gamma$
and $K_L\to\mu^+\mu^- e^+ e^-$. Finally, certain constraints
can be derived from QCD for $f(q^2,q^2)$ in the limit $q^2\gg m^2_V$.
Using this framework, the authors of \cite{DIP} derive
\begin{equation}\label{asddip}
|A_{SD}|^2 < 2.8\cdot 10^{-9}
\end{equation}
which is very similar to (\ref{asdgdp}). 
A recent discussion of $K_L\to\mu^+\mu^-$ can also be found
in \cite{GV}, where the difficulty of extracting
short-distance information from this decay in a truly
model-independent way is particularly emphasized.

\section{$\mu$-Polarization in $K$ Decays}

Measuring the polarization of muons from $K$ decays allows
one to study a number of interesting CP-odd or T-odd
observables. In general such observables are very small
in the Standard Model. However, the expected small effects
are, in several cases, theoretically quite well under control.
Muon polarization observables are then particularly suited
as genuine probes of new interactions and thus provide
us with additional and complementary tools to explore the
physics of flavour.

\begin{figure}
 \vspace{4.5cm}
\includegraphics{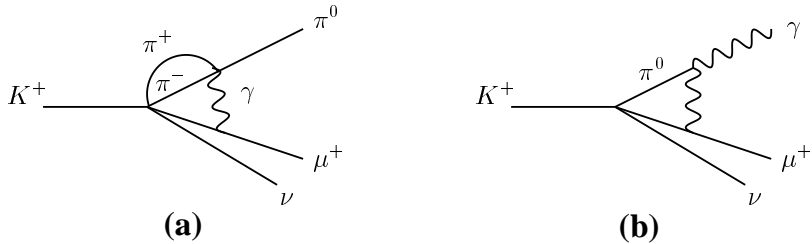}
 \caption{\it
      Final state interactions generating transverse muon
   polarization in kaon decays within the Standard Model:
   (a) $K^+\to\pi^0\mu^+\nu$, (b) $K^+\to\mu^+\nu\gamma$.
    \label{fig:kmup} }
\end{figure}

A typical example is the transverse muon polarization
\begin{equation}\label{ptmu}
P^\mu_T=\langle\hat s_\mu\cdot
\frac{(\vec p_\mu\times \vec p_\pi)}{|\vec p_\mu\times \vec p_\pi|}
\rangle
\end{equation}
in $K^+\to\pi^0\mu^+\nu$ decay. A nonvanishing polarization
could arise from the interference of the leading, standard
W-exchange amplitude with a charged-higgs exchange contribution
involving CP violating couplings. $P^\mu_T$ is therefore an interesting
probe of New Physics \cite{GK,FV} with conceivable effects
of up to $P^\mu_T\sim 10^{-3}$. 
Planned experiments could reach a sensitivity
of $P^\mu_T\sim 10^{-4}$ \cite{ABE,DIW}.
Independently of CP or T violation a nonvanishing $P^\mu_T$ can in
principle be induced by final state interactions (FSI).
Note that $P^\mu_T\not= 0$ is not forbidden by CP or T symmetry,
although it can be induced when these symmetries are violated.
In the case of $K^+\to\pi^0\mu^+\nu$ FSI phases arise only at
two loops in QED (fig. \ref{fig:kmup} (a)) and are very small 
($P^\mu_T(FSI)\sim 10^{-6}$) \cite{ZHI}.
This peculiar feature of  $K^+\to\pi^0\mu^+\nu$ is to be contrasted
with the case of $K^0\to\pi^-\mu^+\nu$, where the
final state contains two charged particles. Correspondingly
the FSI phase, now a one-loop effect, is then much larger.

Transverse muon polarization may also be studied using the
radiative decay $K^+\to\mu^+\nu\gamma$.
Also in this case the Standard Model effect from
electromagnetic final state interactions is generated already
at one loop (fig. \ref{fig:kmup} (b))
and therefore more prominent than in
$K^+\to\pi^0\mu^+\nu$. A recent detailed study of this
mechanism has been performed in \cite{HI}. It is found
that transverse polarization can be computed quite reliably
and occurs at the level of $P^\mu_T(FSI)\sim 10^{-4}$.
Concerning potential signals of New Physics, \cite{HI}
estimate that $P^\mu_T \stackrel{<}{_\sim} 10^{-4}$ in
generic supersymmetric models with unbroken R-parity.
Larger effects could arise if R-parity is broken \cite{CGL} 
or with an extended Higgs sector \cite{GLKLO}.
In the absence of an enhancement by New Physics, transverse
muon polarization in $K^+\to\mu^+\nu\gamma$, observed at
the $10^{-4}$ Standard Model level, could still be a valuable
experimental cross-check on an eventual signal in
$K^+\to\pi^0\mu^+\nu$, as both decays can be studied with the
same apparatus \cite{RA}.

Another interesting effect is the CP violating longitudinal
muon polarization asymmetry $P^\mu_L$ in $K_L\to\mu^+\mu^-$
decay \cite{EP}. As discussed in the previous section, it is very
difficult to compute theoretically the branching fraction for this
mode. However, using the measured branching ratio as an input,
$P^\mu_L$ can be rather reliably calculated in the Standard Model
using chiral perturbation theory. This is because the dominant
effect, from indirect CP violation related to $\varepsilon_K$,
proceeds through the $K_S\to\gamma^*\gamma^*\to\mu^+\mu^-$
amplitude, which is well under control in the chiral
perturbation theory framework. The calculation gives
$P^\mu_L\approx 2\cdot 10^{-3}$ \cite{EP}. An effect significantly
above this level would be a clear signal of New Physics.
Unfortunately, the $K_L\to\mu^+\mu^-$ branching ratio is
very small and an experiment to measure $P^\mu_L$ appears
challenging.

\section{Summary}

In this talk we have reviewed a selection of important
topics in the field of rare kaon decays, highlighting in
particular some recent developments.

The rare decay modes $K^+\to\pi^+\nu\bar\nu$ and, even more so,
$K_L\to\pi^0\nu\bar\nu$ clearly stand out as excellent, theoretically
clean probes of both the standard theory of flavour and the physics
beyond it.
Similar physics can be addressed by studying $K_L\to\pi^0e^+e^-$,
although the theoretical situation is considerably more complex
in this case. An experimental study of the related decay
$K_S\to\pi^0e^+e^-$ will be needed as crucial input for a better
understanding of CP violation in $K_L\to\pi^0e^+e^-$. The required
measurements could possibly be performed at KLOE (Frascati) and
NA48 (CERN).

The process $K_L\to\mu^+\mu^-$ is already accurately measured today.
On the other hand it continues to present a big challenge for theory.
Recent theoretical efforts may lead to an improved understanding
of the long-distance dynamics that determines this mode.

Muon polarization observables in $K^+\to\pi^0\mu^+\nu$,
$K^+\to\mu^+\nu\gamma$ or $K_L\to\mu^+\mu^-$ provide
interesting tests of the flavour sector, complementary to
measurements of rare decay branching fractions. In the Standard
Model these muon polarization effects are very small, which is 
assured in general with good theoretical reliability. They qualify
therefore as genuine probes of New Physics.

There are many further possibilities, e.g. lepton flavour
violating modes ($K_L\to\mu e$, $K\to\pi\mu e$) probing short-distance
physics, or, on the other hand, studies of chiral perturbation
theory to describe long-distance dominated kaon modes. The latter
are of great interest not only as tests of strong interaction physics
in low-energy weak processes from first principles, but also as
a framework for assessing the effects of long-distance dynamics
on the extraction of short-distance flavour physics.

The broad variety of possibilities and the compelling physics
motivation pro\-mise an exciting long term future for studies of
rare K decays.

\section*{Acknowledgements}

I thank the organizers of DA$\Phi$NE '99 for the invitation to
this very pleasant and informative meeting and in particular
Gino Isidori for the kind hospitality at Frascati.

\end{document}